\documentclass[aps,prd,preprint,tightenlines,groupedaddress,showpacs]{revtex4}
\usepackage{epsf,epsfig,graphics,graphicx}
\begin{document}

\title{A numerical study of pseudoscalar inflation with an axion-gauge field coupling}

\author{Shu-Lin Cheng$^1$}
\author{Wolung Lee$^1$}
\author{Kin-Wang Ng$^{2,3}$}

\affiliation{
$^1$Department of Physics, National Taiwan Normal University,
Taipei 11677, Taiwan\\
$^2$Institute of Physics, Academia Sinica, Taipei 11529, Taiwan\\
$^3$Institute of Astronomy and Astrophysics, Academia Sinica, Taipei 11529, Taiwan}

\date{\today}

\begin{abstract}
A numerical study of a pseudoscalar inflation having an axion-photon-like coupling is performed by solving numerically the coupled differential equations of motion for inflaton and photon mode functions from the onset of inflation to the end of reheating. The backreaction due to particle production is also included self-consistently. We find that this particular inflation model realizes the idea of a warm inflation in which a steady thermal bath is established by the particle production. In most cases this thermal bath exceeds the amount of radiation released in the reheating process. In the strong coupling regime, the transition from the inflationary to the radiation-dominated phase does not involve either a preheating or reheating process. In addition, energy density peaks produced near the end of inflation may lead to the formation of primordial black holes.
\end{abstract}

\maketitle

\section{Introduction}

The inflation model, in which our observable
Universe is only a tiny local patch of a causally connected
region that underwent an exponential or de Sitter expansion at early times
driven by an inflaton potential, is generally accepted for
explaining the observed spatially flat and homogeneous Universe.
A simple version of the model such as the slow-roll inflation predicts quasi de Sitter
vacuum fluctuations during inflation which could give rise to Gaussian and
nearly scale-free metric perturbation containing both matter density fluctuations
(scalar mode) and gravitational waves (tensor mode)~\cite{olive}.

Recent astrophysical and cosmological observations such as cosmic microwave background anisotropies,
gravitational lensing, baryon acoustic oscillations, supernovae, and Hubble constant have supported
the slow-roll scenario, and further constrained the non-Gaussianity of the matter perturbation and the
tensor-to-scalar ratio~\cite{planck1}. This enables us to discriminate between different types of inflation
or even disfavor certain models~\cite{planck2}. However, some fundamental questions have yet to be answered.
What is the nature of the inflaton?
Do classical matter density inhomogeneities in the present Universe genuinely
come from vacuum fluctuations of the inflaton?
Are the matter density fluctuations truly Gaussian?
Is the tensor-to-scalar ratio predictable?
Future cosmic microwave background measurements and large-scale-structure surveys should be
able to address some of these questions or pose a challenge to the standard slow-roll model.

There has been a lot of studies on inflationary models that go
beyond the simplest single-field, slow-roll inflation. Recently, a class of
models has considered a new source for generating metric
perturbation during the standard slow-roll inflation through
a coupling between the inflaton and other quantum
fields, with particular attentions to the influences of particle production
and its associated backreaction to the slow-roll inflation.
This leads to very interesting effects such as
the suppression of large-scale density fluctuations~\cite{wu}, the
generation of non-Gaussian and non-scale-invariant density power
spectrum~\cite{barnaby,lee,meer}, the damping of primordial
gravitational waves due to a copious production of free-streaming particles~\cite{ng},
the generation of stochastic gravitational waves that
could be detected at ground-based gravity-wave interferometers~\cite{barnaby2,cook},
and the formation of primordial black holes at density peaks near
the end of inflation~\cite{lin,linde,bugaev}.

In particular, a lot of attentions have been paid to the study of the phenomenological effects
on inflation via a pseudoscalar-vector coupling~\cite{anber,anber2,durrer,barnaby,meer,barnaby2,cook,lin,linde,bugaev},
\begin{equation}
\label{int}
  \mathcal{L}_{\mathrm{int}} = - \frac{\alpha}{4 f}\, \varphi \, F^{\mu\nu}\tilde{F}_{\mu\nu}
\end{equation}
where $f$ is an energy scale and $\alpha$ is a dimensionless coupling constant. The pseudoscalar
$\varphi$ is an axion-like field that drives inflation,
$F_{\mu\nu} = \partial_\mu A_\nu - \partial_\nu A_\mu$ is the vector field strength tensor, and
$\tilde{F}^{\mu\nu} = \frac{1}{2}\epsilon^{\mu\nu\alpha\beta} F_{\alpha\beta}$ is its dual.
The pseudoscalar can be a pseudo Nambu-Goldstone boson in the natural inflation~\cite{freese} or the
many of its variant models~\cite{axion}, which is expected to couple to some gauge field
as the one shown in Eq.~(\ref{int}).
In addition, axion inflation is generically at work in a broad class of inflationary models in supergravity,
in which reheating after the end of inflation requires considering the pseudoscalar-vector coupling~\cite{linde}.
So far, tackling the inflaton-vector system has relied on an adiabatic approximation under which simple analytic
solutions for the vector mode equations can be used. Although this approximation is valid in the beginning
of inflation, it becomes unreliable near the end of inflation and even inapplicable in the era of reheating,
when physical effects of main concern are taking place. Recently, the authors in Ref.~\cite{adshead}
have studied using lattice simulations the onset of the reheating epoch at the end of inflation.
They found that for a wide range of parameters preheating is efficient and that
in certain cases the inflaton transfers all its energy to the vector fields within a few inflaton oscillations.
In this paper, we perform a numerical analysis of the system from the onset of inflation towards the end of reheating,
solving self-consistently the coupled equations of motion of the fields including field perturbation and backreaction.
In particular, we pay attention to the transition from the end of inflaton to the preheating. We will review the
the inflation model in Sec.~\ref{model} and set up the numerical scheme in Sec.~\ref{scheme}.
Sec.~\ref{results} is our numerical results that are compared to those in Ref.~\cite{adshead}.
We make our conclusions in Sec.~\ref{conclusion}.

\section{Pseudoscalar inflation}
\label{model}

We consider a simple model of inflation driven by a pseudoscalar that couples to a $U(1)$ gauge field via the interaction~(\ref{int}).
The action is given by
\begin{equation}
\label{action}
  \mathcal{S} =  \int d^4 x \sqrt{-g} \left[ \frac{M_p^2}{2} \, R  -\frac{1}{2}\partial_\mu\varphi \partial^\mu\varphi- V(\varphi) - \frac{1}{4}F^{\mu\nu}F_{\mu\nu} - \frac{\alpha}{4 f} \varphi  \, \tilde{F}^{\mu\nu} \, F_{\mu\nu} \right],
\end{equation}
where $R$ is the curvature scalar, $M_p$ is the reduced Planck mass, $V(\varphi)$ is the inflaton potential,
$F_{\mu \nu} = \partial_\mu A_\nu - \partial_\nu A_\mu$ is the field strength tensor, and
$\tilde{F}^{\mu\nu} = \frac{1}{2}\epsilon^{\mu\nu\alpha\beta} F_{\alpha\beta}/{\sqrt{-g}}$ is its dual.
Note that $1/{\sqrt{-g}}$ is added to the dual tensor because $\epsilon^{\mu\nu\alpha\beta}$ is a tensor density of weight $-1$.
Here we assume a spatially flat Friedmann-Robertson-Walker metric:
\begin{equation}
ds^2=-g_{\mu\nu} dx^\mu dx^\nu= a^2(\eta) (d\eta^2- d \vec{x}^2),
\end{equation}
where $a(\eta)$ is the cosmic scale factor and $\eta$ is the conformal time related to the cosmic time by $dt=a(\eta)d\eta$.
The Hubble parameter $H \equiv \dot{a} / a$ has a conformal time analogue ${\cal H} \equiv a' / a$, where the dot and the prime
denote derivatives with respect to $t$ and $\eta$, respectively.

From the action~(\ref{action}), we can write down the Friedmann equation, the equation of motion for the inflaton, and the Maxwell equations, respectively:
\begin{eqnarray}
&& {\cal H}^2 = \frac{1}{3 M_p^2} \left[ \frac{1}{2} \left(\frac{\partial \varphi}{\partial \eta}\right)^2 + \frac{1}{2} \left( \vec{\nabla}  \varphi \right)^2   + a^2 \, V(\varphi) + \frac{a^2}{2} \left( \vec{E}^2 + \vec{B}^2 \right) \right],\label{G00}\\
&& \frac{\partial ^2\varphi}{\partial \eta^2} + 2 {\cal H} \frac{\partial \varphi}{\partial \eta} - \vec{\nabla}^2 \varphi  + a^2 \, \frac{d V}{d \varphi}  = a^2 \frac{\alpha}{f}  \vec{E}\cdot \vec{B},\label{varphieom}\\
&& \frac{\partial^2 \vec{A}}{\partial \eta^2} - \vec{\nabla}^2 \vec{A}
+ \vec{\nabla} (\vec{\nabla}\cdot \vec{A}) = \frac{\alpha}{f}
\frac{\partial \varphi}{\partial \eta} \vec{\nabla} \times \vec{A} - \frac{\alpha}{f} \vec{\nabla} \varphi \times
\frac{\partial \vec{A}}{\partial \eta},\label{maxwell1}\\
&&\frac{\partial}{\partial \eta}(\vec{\nabla} \cdot \vec{A}) = \frac{\alpha}{f}\vec{\nabla} \varphi \cdot(\vec{\nabla} \times \vec{A}),
\label{maxwell2}
\end{eqnarray}
where for the Maxwell equations we have chosen the temporal gauge, i.e. $A_{\mu} = (0, \vec{A})$, and we have introduced the physical ``electric'' and ``magnetic'' fields,
\begin{equation}
\vec{E} = -{1\over a^2} \frac{\partial \vec{A}}{\partial \eta},\quad \vec{B} = {1\over a^2}\vec{\nabla} \times \vec{A}.
\end{equation}
The above equations form a complete set of coupled differential equations. In the present consideration, we study the production of gauge quanta by the rolling inflaton via the interaction during a slow-roll inflation, taking into account self-consistently the backreaction of the gauge quanta production on inflation. Furthermore, we consider the generation of curvature perturbation in this scenario.

\subsection{Gauge quanta production and backreaction}

To calculate the production of gauge quanta, we separate the inflaton into a mean field and its fluctuations:
\begin{equation}
  \varphi = \phi (t) + \delta \varphi (t ,\, \vec{x}).
\end{equation}
To the first order in quantum fields, we can consistently impose the condition, $\vec{\nabla} \cdot \vec{A} = 0$. Then, Eqs.~(\ref{maxwell1}) and (\ref{maxwell2}) become a modified wave equation,
\begin{equation}
\frac{\partial^2 \vec{A}}{\partial \eta^2} - \vec{\nabla}^2 \vec{A} - \frac{\alpha}{f} \phi' \vec{\nabla} \times \vec{A} =0.
\label{wave}
\end{equation}
To proceed, we decompose the gauge field $\vec{A}(\eta,{\vec x})$ as
\begin{equation}
 \vec{A}(\eta,{\vec x}) = \sum_{\lambda=\pm} \int \frac{d^3k}{(2\pi)^{3/2}} \left[ \vec{\epsilon}_\lambda({\vec k}) a_{\lambda}({\vec k}) A_\lambda(\eta,{\vec k}) e^{i {\vec k}\cdot {\vec x}} + \mathrm{h.c.}   \right],
\label{decomposition}
\end{equation}
where the annihilation and creation operators obey
\begin{equation}
\left[a_{\lambda}({\vec k}), a_{\lambda'}^{\dagger}({\vec k'})\right] = \delta_{\lambda\lambda'}\delta ({\vec k}-{\vec k}'),
\end{equation}
$\vec{\epsilon}_\lambda$ are normalized circular polarization vectors satisfying
$\vec{k}\cdot \vec{\epsilon}_{\pm} \left( \vec{k} \right) = 0$,
$\vec{k} \times \vec{\epsilon}_{\pm} \left( \vec{k} \right) = \mp i k \vec{\epsilon}_{\pm} \left( \vec{k} \right)$,
$\vec{\epsilon}_\pm \left( -\vec{k} \right) = \vec{\epsilon}_\pm \left( \vec{k} \right)^*$, and
$\vec{\epsilon}_\lambda \left( \vec{k} \right)^* \cdot \vec{\epsilon}_{\lambda'} \left( \vec{k} \right) = \delta_{\lambda \lambda'}$.
Inserting the decomposition~(\ref{decomposition}) into Eq.~(\ref{wave}), we obtain the equation of motion for the mode functions,
\begin{equation}
  \left[ \frac{\partial^2}{\partial\eta^2} + k^2 \mp 2aH k\xi\right] A_{\pm}(\eta,k) = 0, \hspace{5mm} \xi \equiv \frac{\alpha \dot{\phi}}{2 f H}\,.
\label{f_V}
\end{equation}
It is well known that either one of the two modes exhibits a spinoidal instability as long as the modes satisfy
the condition, $k/(aH)<2\vert\xi\vert$. When the inflaton rolls down the potential,
these unstable modes grow exponentially by consuming the inflaton kinetic energy. As such,
the energy density of the produced gauge quanta is given by the vacuum expectation value,
\begin{equation}
\frac{1}{2}\langle \vec{E}^2+\vec{B}^2 \rangle = \frac{1}{4 \pi^2 a^4} \int d k \,  k^2 \sum_{\lambda=\pm}\left( \vert A_\lambda' \vert^2 + k^2 \vert A_\lambda \vert^2 \right).
\label{f_Y}
\end{equation}
Also, we have
\begin{equation}
\langle \vec{E} \cdot \vec{B} \rangle = - \frac{1}{4 \pi^2 a^4} \int d k \, k^3 \, \frac{d}{d \eta} \left(\vert A_+ \vert^2-\vert A_- \vert^2\right).
\label{f_X}
\end{equation}
On the other hand, the production of gauge quanta gives rise to a backreaction on the background. The background evolution is therefore governed by the mean field parts of Eqs.~(\ref{G00}) and (\ref{varphieom}) with the backreaction effects included, i..e
\begin{eqnarray}
\label{f_bg}
  &&  \ddot{\phi} + 3 H \dot{\phi} +\frac{dV}{d\phi} = \frac{\alpha}{f} \langle \vec{E}\cdot \vec{B} \rangle, \\
\label{f_H}
  && 3 H^2 = \frac{1}{M_p^2} \left[ \frac{1}{2}\dot{\phi}^2 + V(\phi) + \frac{1}{2} \langle \vec{E}^2 + \vec{B}^2 \rangle \right].
\end{eqnarray}

\subsection{Inflaton fluctuations and curvature perturbation}
\label{curpert}

To study the generation of inflaton fluctuations during inflation, one can integrate out the gauge fields in the real-time formalism to obtain an effective equation of motion for the inflaton fluctuations, including backreaction and dissipation effects. Similar work about integrating out scalar fields has been done in Refs.~\cite{wu,lee}. However, in Refs.~\cite{anber2,barnaby2} it was proposed that one can estimate these effects in the regime of strong backreaction by incorporating a dissipation term in the equation of motion:
\begin{equation}
  \left[ \frac{\partial^2}{\partial t^2} +3 \beta H \frac{\partial}{\partial t} - \frac{{\vec\nabla}^2}{a^2} + \frac{d^2V}{d\phi^2} \right] \delta\varphi(t,{\vec x})
  = \frac{\alpha}{f} \, \left(  \vec{E}\cdot\vec{B} - \langle \vec{E}\cdot\vec{B} \rangle \right),
\label{perteqn}
\end{equation}
where
\begin{equation}
\beta\equiv 1-2\pi\xi \frac{\alpha}{f} \frac{\langle \vec{E}\cdot\vec{B} \rangle}{3H\dot{\phi}}.
\label{betaequiv}
\end{equation}
Furthermore, it was shown that the solution of this equation can be well approximated  by~\cite{barnaby2,linde}
\begin{equation}
\delta\varphi= \frac{\alpha\left(  \vec{E}\cdot\vec{B} - \langle \vec{E}\cdot\vec{B} \rangle \right)}{3\beta f H^2},
\end{equation}
which leads to a contribution to the power spectrum of the curvature perturbation given by
\begin{equation}
\Delta_\zeta^2(k)=\langle\zeta(x)^2\rangle=\frac{H^2}{\dot\phi^2}\langle\delta\varphi^2\rangle
=\left(\frac{\alpha\langle \vec{E}\cdot\vec{B} \rangle}{3\beta f H\dot\phi}\right)^2.
\label{f_powersp}
\end{equation}
When the backreaction becomes large, the second term in Eq.~(\ref{betaequiv}) dominates and we can have approximately
\begin{equation}
\Delta_\zeta^2(k)\simeq \left(\frac{1}{2\pi\xi}\right)^2,
\quad
\langle\delta\varphi^2\rangle\simeq \left(\frac{f}{\pi\alpha}\right)^2,
\label{largealpha}
\end{equation}

\section{Numerical scheme}
\label{scheme}

We employ the chaotic inflation scenario~\cite{chaotic} to unravel the particle production and its backreaction by full numerical calculations. In such a universe, the standard slow-roll inflationary background is driven by the potential
\begin{equation}
V(\varphi)={m^2\over 2}\varphi^2,
\end{equation}
where the parameter $m=1.8\times 10^{13}$ GeV characterizes the mass of the inflaton field. Here we rescale all dynamical variables in terms of the reduced Planck mass, $M_p=2.435\times 10^{18}$ GeV. Hence, $m=7.39\times 10^{-6}$. We set $f=1$ and $\phi_0=-14.15$ and $\dot\phi_0=6\times 10^{-6}$ respectively for the initial position and the initial velocity of the inflaton $\varphi$. The e-folding since the beginning of inflation is defined by $N(t)=\int_0^t H(t') dt'$, where it is chosen that $a_0=1$.

\begin{figure}[htp]
\centering
\includegraphics[width=0.8\textwidth]{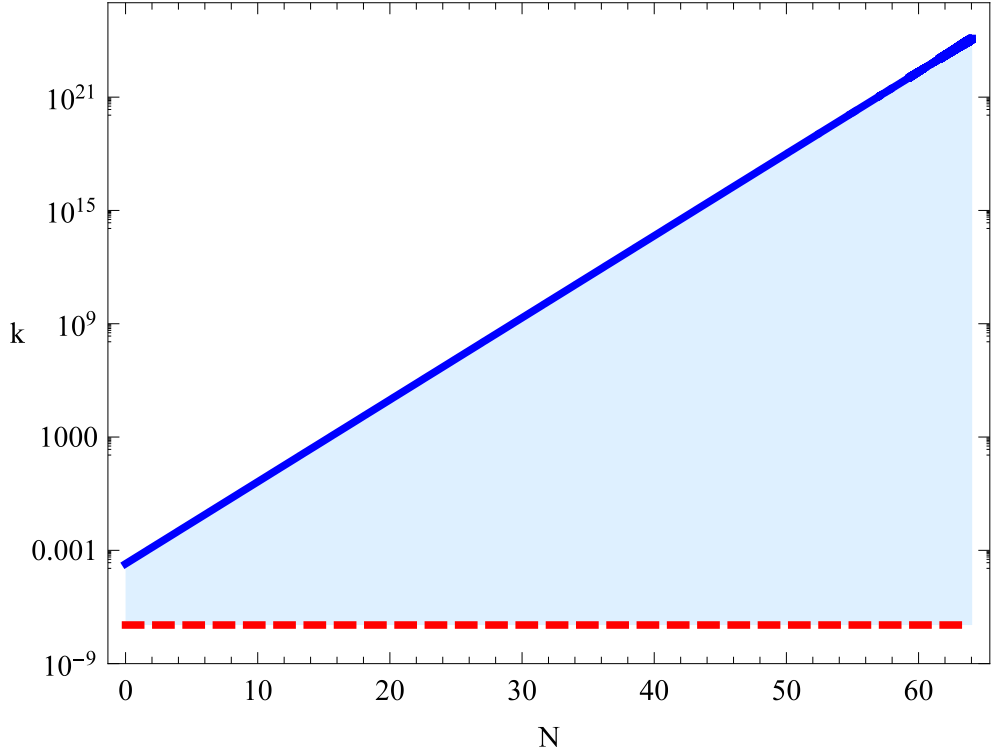}
\caption{Solid line is the upper limit and the dashed line is the lower limit. The shaded area between the two limits represents the range of $k$-modes at a given e-folding $N$ that contribute to the backreaction. $k$ in this figure and all dynamical variables in the following figures are
rescaled by the reduced Planck mass $M_p=2.435\times 10^{18}$ GeV. }
\label{p_qrange}
\end{figure}

The background evolution $\phi(t)$ and $a(t)$ with feedback from gauge quanta production can be obtained numerically by solving self-consistently Eqs.~(\ref{f_Y}), (\ref{f_X}), (\ref{f_bg}), and (\ref{f_H}), where the mode functions of the gauge field during inflation are governed by Eq.~(\ref{f_V}), with the initial conditions of the gauge mode functions given by $ A_{\pm 0} = 1/\sqrt{2k}$ and $\dot A_{\pm 0} = -ik/\sqrt{2k}$.
To solve this set of coupled differential equations, we use the second-order Runge-Kutta method. To accurately capture the feedback of the gauge quanta production in Eqs.~(\ref{f_bg}) and (\ref{f_H}), we need to integrate over the range of $k$ to which the $A_{\pm}$ modes have contributions at a given time. Initially, all $A_{\pm}$ $k$-modes have small contributions. However, the $A_+$ modes develop spinoidal instability and grow exponentially when they satisfy the condition $k/(aH)<2\xi$. On the other hand, the backreaction of the $k$-modes decay with the growth of $a$, so one should only count $A_{\pm}$ modes that satisfy $(8\xi)^{-1}<k/(aH)<2\xi$~\cite{barnaby}. Note that the range is time varying during inflation. However, we include all $k$-modes with $10^{-4}<k$. The scale factor $a$ will grow up to about $e^{60}$ at the end of inflation, so we have to calculate a very large range of
$k$-modes ($10^{-4}<k<10^{23}$). Since it is difficult to calculate very high-$k$ modes that are varying too fast with time, to reduce the computing load we approximate these modes by their initial values until they enter the range with $k/(aH)<2\xi$. Although this would cause phase shifts of the mode functions, it can hardly affect the time evolution of the mode amplitudes. Since only the mode amplitudes are used in the background equations, the approximation works very well as far as the background evolution is concerned. Figure~\ref{p_qrange} shows the range of $k$-modes that are included in the backreaction at a given time during inflation.

To calculate the backreaction, we need to integrate all relevant $k$-modes by the Trapezoid rule, the Simpson's rule, and/or the Boole's rule. In practice, because of the range of $k$ relevant to the backreaction is too big, it is problematic for us just to divide the whole range into equal integration intervals. As such, we do the following trick. Initially, the range of $k$ is narrow and the value of $k$ is small. During inflation, the range of $k$ becomes larger and larger, so we choose different partitions for different $k$-modes. The partition in low-$k$ region is small whereas that in high-$k$ region is wide. In practice, we adopt equal partitions in log-$k$ range. By integrating the modes in each interval and then summing all intervals we thus obtain the total backreaction effect to the background evolution. Furthermore, we will increase the time steps as well as the $k$-resolution until we obtain steady results.

Because of limited computational resource, in some cases we have tested our concluding results in toy inflation models that have shortened duration of inflation while capturing relevant features of the particle production and the backreaction to the background evolution. In this toy inflation model, the range of $k$ is considerably reduced and therefore we can save more computational power to increase the time steps, $k$-resolution, and $k$-range for calculating the growth of the $k$-modes and their feedback. We have found that this strategy is very useful to eliminate computational artifacts and ascertain the physical results obtained in realistic inflation models.

\section{Numerical results}
\label{results}

\begin{figure}[htp]
\centering
\includegraphics[width=0.8\textwidth]{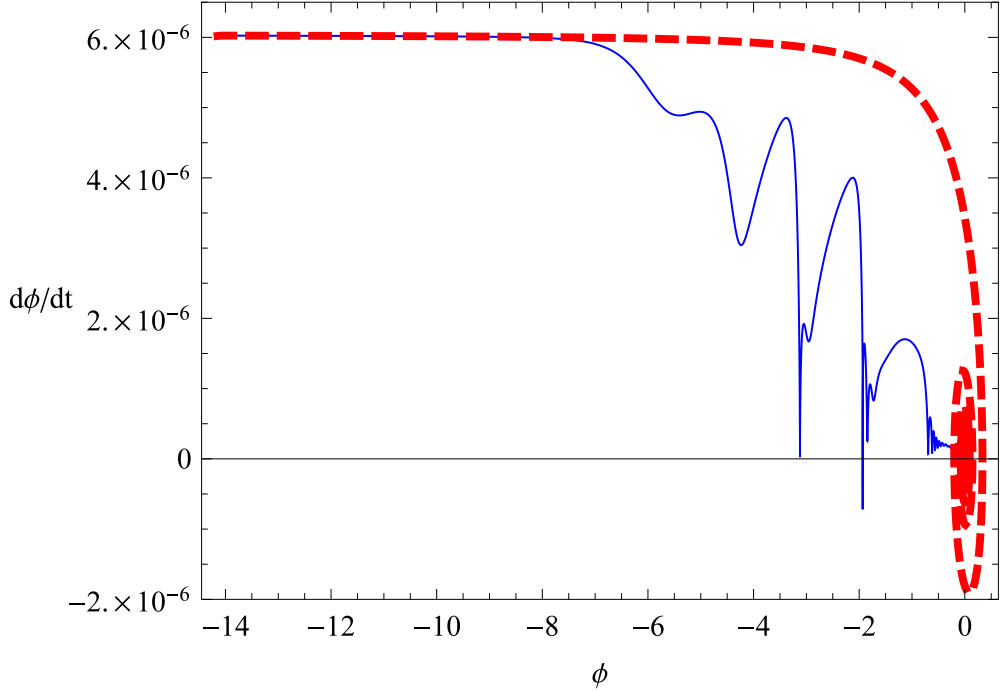}
\caption{Phase diagram of the inflaton mean field $\phi$ and its velocity $\dot\phi$, with the initial $\phi_0=-14.15$, $\dot\phi_0=6\times10^{-6}$, and the coupling constant $\alpha=32$. This case corresponds to $\xi=2.2$ in the beginning of inflation ($N=0$). The solid line does take the backreaction into account, while the dashed line does not.}
\label{p_phase}
\end{figure}

\begin{figure}[htp]
\centering
\includegraphics[width=0.8\textwidth]{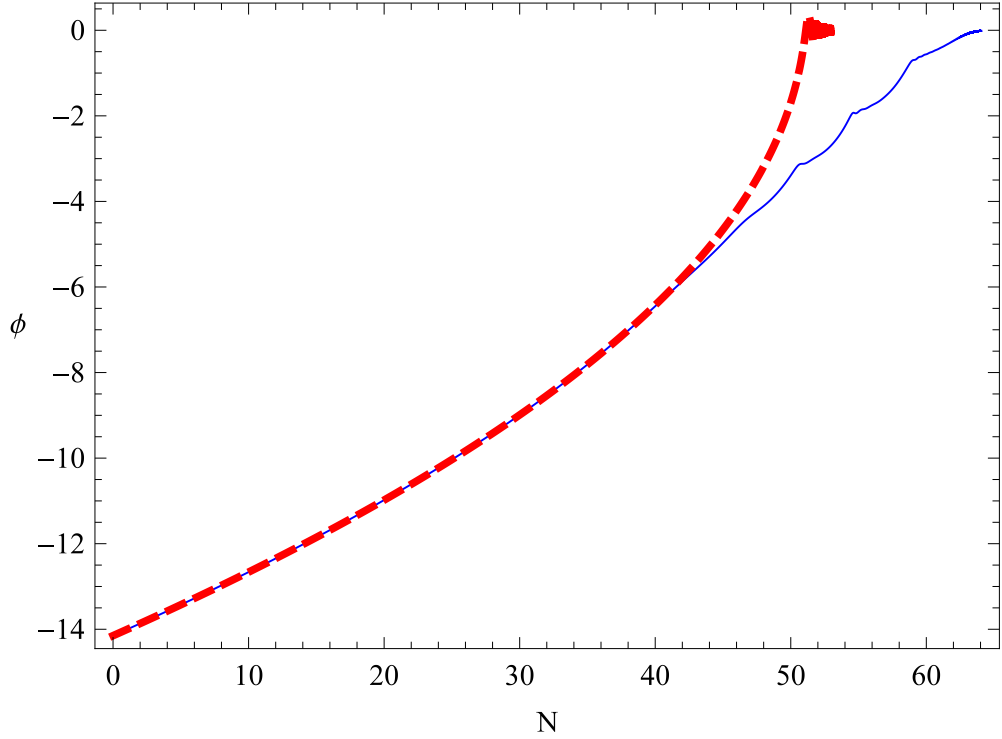}
\caption{Evolution of $\phi$ as a function of the e-folding $N$ in the case with $\alpha=32$. The solid line does take the backreaction into account, while the dashed line does not.}
\label{p_phi}
\end{figure}

\begin{figure}[htp]
\centering
\includegraphics[width=0.8\textwidth]{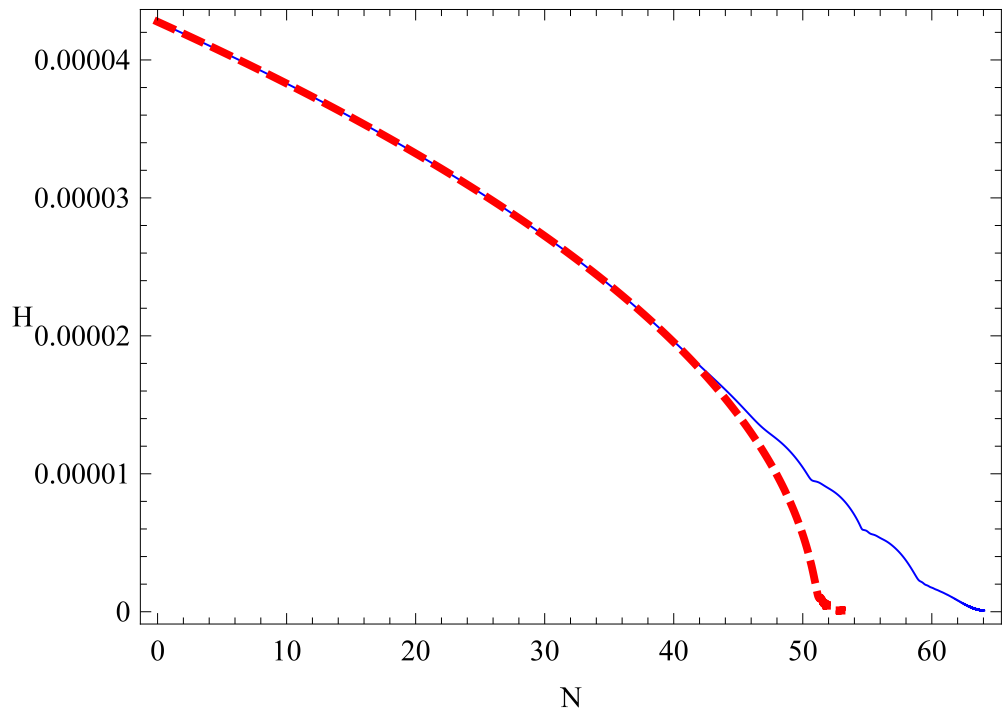}
\caption{Time evolution of the Hubble parameter $H$ in the case with $\alpha=32$. The solid line does take the backreaction into account, while the dashed line does not.}
\label{p_H}
\end{figure}

\begin{figure}[htp]
\centering
\includegraphics[width=0.8\textwidth]{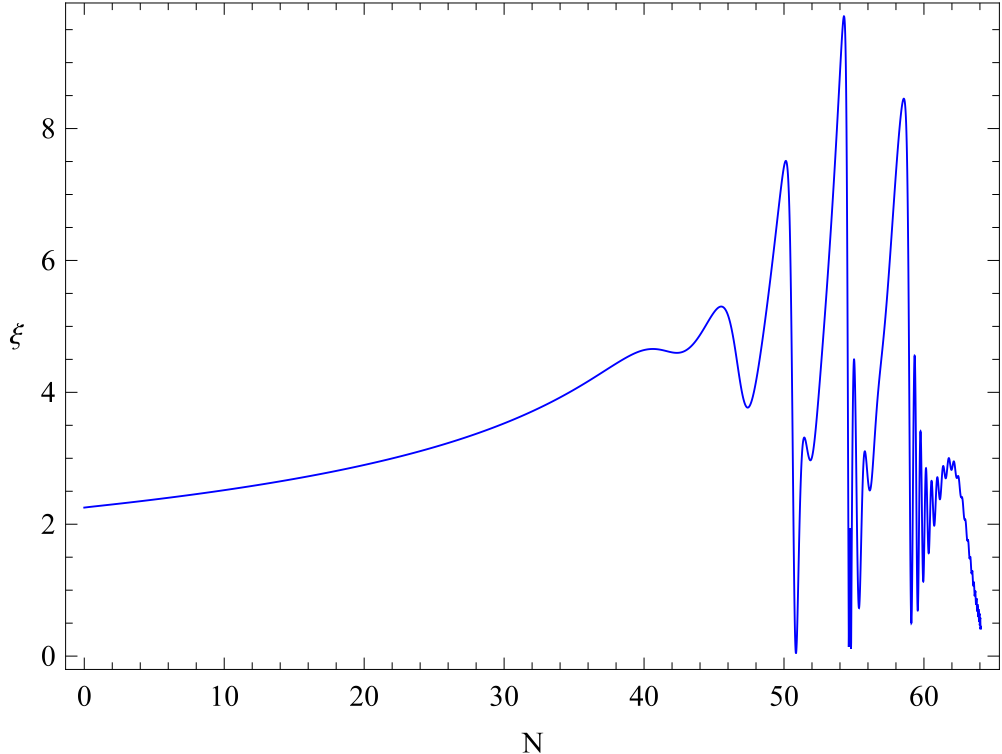}
\caption{Time evolution of the parameter $\xi$ in Eq.~(\ref{f_V}) in the case with $\alpha=32$.}
\label{p_xi}
\end{figure}

Let us first consider a strong coupling $\alpha=32$. In this case, $\xi=2.2$ in the beginning of inflation at $N=0$ and inflation ends at $N=64$. This value of $\xi$ is the upper bound coming from measurements of the power spectrum and the bispectrum of the cosmic microwave background~\cite{meer}.
The influence of the backreaction of the gauge quanta production to the inflaton motion is shown in Fig.~(\ref{p_phase}). Because the range of $k$-modes for the growing gauge field is related to $\dot\phi$ that in turn affects the backreaction, the temporal variation of $\dot\phi$ is very complicated during inflation. This is the reason that the scalar field $\phi$ and the Hubble parameter $H$ exhibit bouncing at the late stage of inflation. We have also confirmed the occurrence of this bouncing behaviour in a toy inflation model with the strategy as mentioned in Sec.~\ref{scheme}. The numerical results for the evolution of $\phi$ and $H$ are shown in Fig.~\ref{p_phi} and Fig.~\ref{p_H}, respectively. It is apparent that the backreaction slows down the inflaton motion and thus prolong the duration of inflation by about $10$ e-foldings. It is also interesting to see that inflation ends gracefully without preheating or reheating process. These results are consistent with those found in Ref.~\cite{linde}. Here we show detailed features of the evolution towards the end of inflation. Figure~\ref{p_xi} shows the evolution of the parameter $\xi$. The fast oscillatory behaviour of $\xi$ at the late stage of inflation puts a caution that one cannot use the analytic solution for the growing mode $A_+$, which has assumed an adiabatic condition that treats $\xi$ as a constant parameter (see Ref.~\cite{linde} and therein previous work), for the whole course of inflation, especially near the end of inflation.

\begin{figure}[htp]
\centering
\includegraphics[width=.9\textwidth]{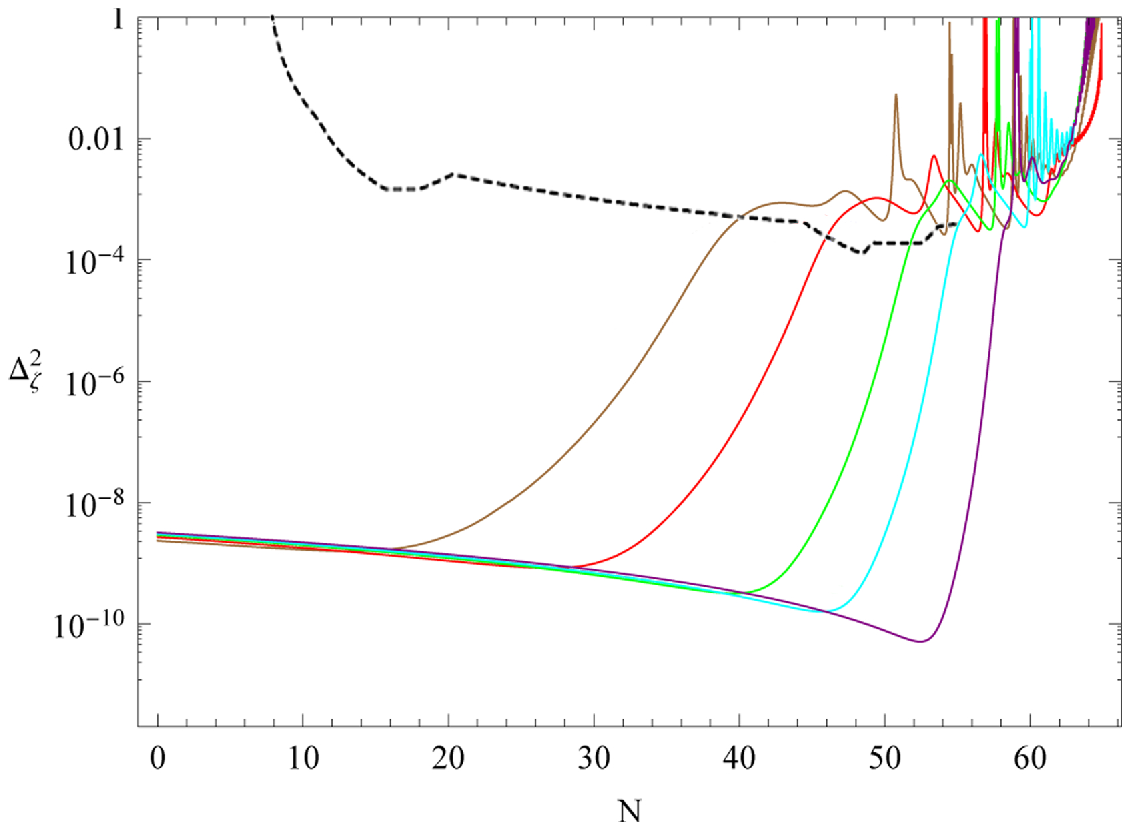}
\caption{Time evolution of the power spectra in Eq.~(\ref{f_powersp}) during inflation. The lines shown in the figure, in order from left to right, are for $\alpha=32, 28, 23, 20, 16$, respectively. These correspond in turn to $\xi(N=0)=2.2, 1.91, 1.53, 1.32, 1.04$. Note that we have adjusted the value of the initial $\phi_0$ in each case such that the duration of inflation is $64$ e-foldings. The black hole bound is the upper dashed line extended to about $N=54$.}
\label{p_powersp}
\end{figure}

The results of the power spectra of the curvature perturbation in Eq.~(\ref{f_powersp}) for various $\alpha$ parameters are plotted in Fig.~\ref{p_powersp}, where each power spectrum has already included the amplitude of the vacuum inflationary perturbation as predicted in a standard slow-roll inflation, which is given by $H^4/(4\pi^2{\dot\phi}^2)$. Note that we have adjusted the value of the initial $\phi_0$ in each case such that the duration of inflation is $64$ e-foldings. 
In general the results in the figure are consistent with the large coupling approximation for $\Delta_\zeta^2$~(\ref{largealpha}).
The effect of the backreaction can modify drastically the power spectra found in Ref.~\cite{linde} near the end of inflation.
The dashed line in Fig.~\ref{p_powersp} is the upper bound on the power spectrum coming from the non-detection of primordial black holes (referring to Fig.~5 of Ref.~\cite{linde} and noting that the bound ends at about $N=54$). We find that the black hole bound constrains $\alpha<20$, which is consistent with, though slightly tighter than, the findings in Refs.~\cite{linde,bugaev}. For $\alpha>20$, the power spectra show prominent peaks that exceed the black hole bound. Interestingly, we observe that near the end of inflation all power spectra show high spikes that seem to be beyond the black hole bound when presumably extrapolated to $N>54$. Note that in some of the
curves, the power spectrum spikes above unity and is cut off there. We warn that in the present consideration the power spectrum is no longer valid when its value becomes near unity. In this situation, one must take into account the effects of gravitation, which could presumably dampen the spikes to a consistent level~\cite{fugita}. 
However, the occurrence of these density spikes suggests that primordial black holes may probably be formed at high-density regions near the end of inflation. If so, a more stringent bound on the value of $\alpha$ can be derived. However, more studies about the formation of black holes at the density spikes in the present consideration are needed to confirm this suggestion.

In Fig.~\ref{p_phase}, the dashed line is the result without backreaction. When the term $(\alpha/f) \langle \vec{E}\cdot \vec{B} \rangle$ on the right-hand side of Eq.~(\ref{f_bg}) is negligible, $\ddot\phi \simeq 0$. The slow-roll condition that $\dot\phi$ keeps steady until the end of inflation is satisfied.  Because the initial conditions are $\phi_0<0$ and $\dot\phi_0>0$, the backreaction $(\alpha/f) \langle \vec{E}\cdot \vec{B} \rangle < 0$. The backreaction kicks in when either one of $A_\pm (k)$ grows exponentially for $k$ satisfying the growth condition $k< a\dot\phi\alpha/f$. Once the backreaction starts to affect the motion of the inflaton, the inflaton decelerates, i.e. $\ddot\phi < 0$. Consequently, $\dot\phi$ is reduced and therefore the largest growing mode of $A_\pm (k)$ in the backreaction stops growing. Then, the backreaction $(\alpha/f) \langle \vec{E}\cdot \vec{B} \rangle$ of the produced gauge quanta will be redshifted as $a^{-4}$, thus allowing $\dot\phi$ to build up itself and letting more $k$-modes grow exponentially in the backreaction to affect the inflaton motion again. Since the backreaction effect of the gauge modes is not immediate, even if a mode starts to affect the inflaton motion, the $k$ value of the mode could be much smaller than the growth threshold, $a\dot\phi\alpha/f$. It will then keep exponentially growing until $\dot\phi$ is reduced below the growth condition. As such, there occurs a delay on the response of the backreaction to $\dot\phi$. This explains why the power spectra show spikes in Fig.~\ref{p_powersp}.

\begin{figure}[htp]
\centering
\includegraphics[width=.8\textwidth]{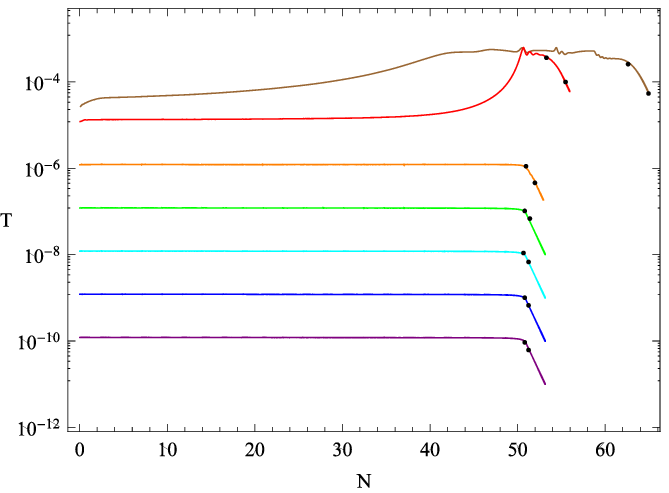}
\caption{Time evolution of the gauge quanta temperature $T$ defined by $\langle \vec{E}^2 + \vec{B}^2 \rangle/2  \equiv (\pi^2/30) g_{\rm eff} T^4$, where $g_{\rm eff} =106.75$. The lines shown in the figure, in order from top to bottom, are for $\alpha=32, 14.5, 1.45, 0.145, 1.45\times 10^{-2}, 1.45\times 10^{-3}, 1.45\times 10^{-4}$, respectively. The left black dot on each line indicates the end of inflation timed by $-\dot{H}/H^2=1$. The right black dot denotes the beginning of the radiation dominated era, after which $T\propto a^{-1}$.}
\label{p_temp}
\end{figure}

Now we consider the reheating temperature of the pseudoscalar inflation. The coupling~(\ref{int}) implies that an inflaton would decay into two gauge bosons.  The perturbative decay rate, i.e. when the mean field $\phi=0$ or the background is the ground state, is given by~\cite{pdg}
\begin{equation}
\Gamma=\frac{\alpha^2 m^3}{64\pi f^2}.
\end{equation}
When the inflaton condensate decays, the Universe is reheated with the gauge quanta.
The gauge quanta may not be thermalized; however, we treat them as a thermal bath with a temperature defined by:
\begin{equation}
\frac{1}{2} \langle \vec{E}^2 + \vec{B}^2 \rangle  \equiv \frac{\pi^2}{30} g_{\rm eff} T^4,
\label{temp}
\end{equation}
where $g_{\rm eff}$ is the number of effectively relativistic degrees of freedom at the time of reheating. Here we assume $g_{\rm eff} = 106.75$.
This perturbative reheating temperature can be estimated by comparing the decay rate with the Hubble rate, $\Gamma \sim 3H$:
\begin{equation}
T_{\rm per}\sim \frac{\alpha}{8\pi}\left(\frac{10}{g_{\rm eff}}\right)^{1/4}\left(\frac{m}{M_p}\right)^{3/2} \left(\frac{M_p}{f}\right) M_p
\sim 4.42\times 10^{-10} \alpha.
\label{Tper}
\end{equation}
However, the spinoidal instability of the gauge field in the presence of the background inflaton mean field may lead to a striking reheating process. Figure~\ref{p_temp} shows the temperature of the gauge quanta in Eq.~(\ref{temp}) from the beginning of inflation to the time of reheating.
Apparently, a higher value of the coupling constant $\alpha$ would lead to a higher reheating temperature.
In the case of $\alpha=32$, because of strong backreaction, in general $\dot{\phi}$ is smaller than that in the standard slow-roll inflation, thus resulting in a longer duration of inflation. There is not a preheating phase; namely, $\phi$ gradually reduces to zero, rather than undergoing a period of field oscillations.
For small coupling constants with $\alpha < 1.45$, the backreaction is no longer important because of the small growth of gauge mode functions for all $k$'s. In these cases, to evaluate the $k$-integration in Eq.~(\ref{f_Y}), we can approximate the mode functions by their initial amplitudes,
$ A_{\pm} \sim 1/\sqrt{2k}$ and $A'_{\pm} \sim \sqrt{k/2}$, and integrate over the range $0<k<2\xi aH$.
Hence, we obtain $\langle \vec{E}^2 + \vec{B}^2 \rangle/2 \sim 2\xi^4 H^4/\pi^2$.
Then, the reheating temperature can be estimated by
\begin{equation}
T_{\rm re} \sim 0.14 \alpha\dot{\phi}/f \sim 0.84\times 10^{-6} \alpha.
\label{Tre}
\end{equation}
This is consistent with the numerical calculations of the reheating temperatures for $\alpha < 1.45$ in Fig.~\ref{p_temp}.
Note that the production of the gauge quanta during inflation can sustain a nearly steady temperature~(\ref{Tre}) which is much higher than the perturbative reheating temperature~(\ref{Tper}). The pseudoscalar inflation with an axion-photon-like coupling is a concrete example that realizes the idea of the existence of a thermal bath during inflation in the so-called warm inflation~\cite{warm}. In most warm inflation models an {\it ad hoc} frictional term ($\Gamma \dot{\varphi}$) reflecting the decay of the inflaton is added to the inflaton equation of motion and the reheating process is through the perturbative decay of inflatons. Also see, for examples in Ref.~\cite{bastero}, some effort to derive the frictional term from first principles.

In the present work, we have solved the coupled differential equations numerically with backreaction included self-consistently. 
Our numerical method is quite similar to that adopted in a previous paper that considered the same axion inflation model but in a different context of magnetogenesis~\cite{fugita}.
In the paper, the authors pointed out that for strong couplings the generation of inflaton fluctuations $\delta\varphi$ by the gauge quanta would become too large to invalidate the linear approximation, and therefore they put forth a consistency condition for linearity, $m^2\langle\delta\varphi^2\rangle<\rho_\phi$, where $\rho_\phi\equiv \dot{\phi}^2/2 + V(\phi)$ is the inflaton energy density. Hence, they found that $\alpha<8.4$ and that for larger $\alpha$ the non-linear coupling between $\delta\varphi$ and the gauge quanta would require performing full lattice simulations. 

However, in Ref.~\cite{fugita}, the inflaton perturbation equation used to calculate the inflaton fluctuations sourced by the gauge quanta does not include a dissipation term. This dissipation term becomes significant in the strong coupling regime, as we have discussed in Sec.~\ref{curpert}. When $\alpha$ becomes large, the inflaton fluctuations $\delta\varphi$ in Eq.~(\ref{perteqn}) are damped by dissipation and their variance is given by $\langle\delta\varphi^2\rangle=f^2/(\pi\alpha)^2$ as shown in Eq.~(\ref{largealpha}). Moreover, the inflaton fluctuations are sourced by the gauge quanta that come from the inflaton energy. Therefore, we propose to improve the consistency condition to that $m^2\langle\delta\varphi^2\rangle<\rho_{\rm tot}$, where $\rho_{\rm tot}\equiv 3M_p^2 H^2$ is the total energy density. Let us denote the energy density of gauge quanta by $\rho_{\rm EM}\equiv \langle \vec{E}^2 + \vec{B}^2 \rangle/2$. Then, we have $\rho_{\rm tot}=\rho_\phi+\rho_{\rm EM}$. During inflation, $\rho_{\rm tot}\simeq \rho_\phi$ and the condition is reduced to $m^2\langle\delta\varphi^2\rangle<\rho_\phi$. During reheating, $\delta\varphi$ can be generated by the gauge quanta even though $\rho_\phi \ll \rho_{\rm EM}$ and the condition becomes $m^2\langle\delta\varphi^2\rangle<\rho_{\rm EM}$. 

\begin{figure}[htp]
\centering
\includegraphics[width=.9\textwidth]{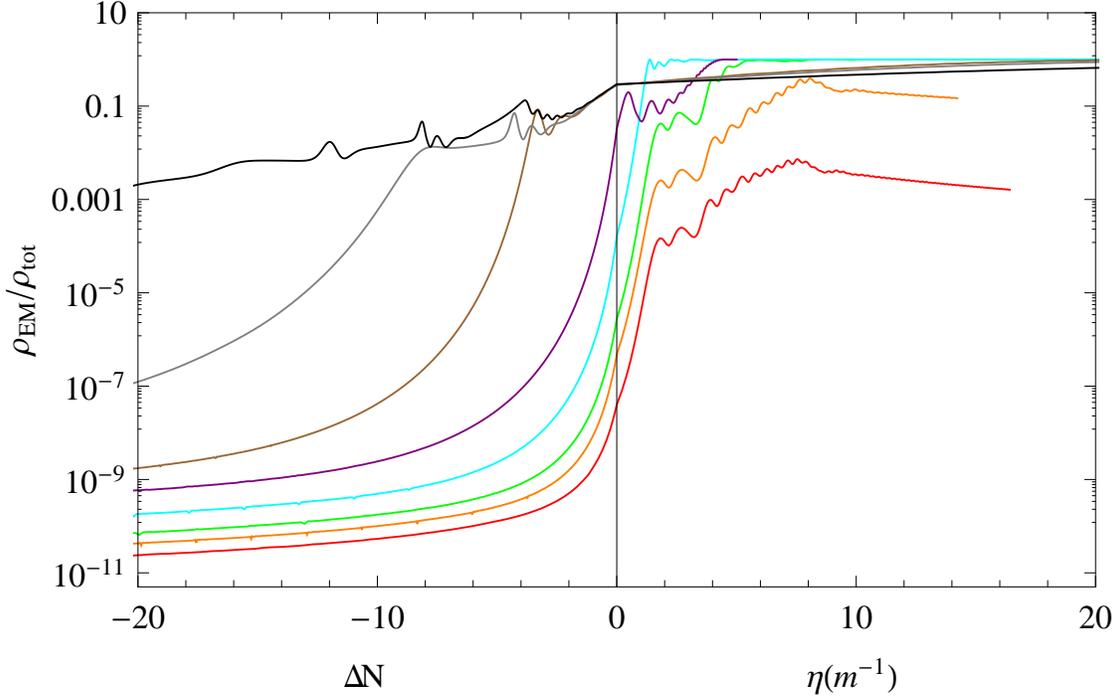}
\caption{Time evolution of the ratio of the energy density of gauge quanta, $\rho_{\rm EM}\equiv \langle \vec{E}^2 + \vec{B}^2 \rangle/2$, and the total energy density, $\rho_{\rm tot}\equiv 3M_p^2 H^2$. Inflation ends at $\Delta N=0$ or $\eta=0$. Before inflation the time is reckoned by how many efoldings prior to the end of inflation. After inflation the scale factor is rescaled to unity and the conformal time $\eta$ is in units of $m^{-1}$. The lines shown in the figure, in order from bottom to top, are for $\alpha=7,8,9,11,14,16,23,32$, respectively.}
\label{EMratio}
\end{figure}

Now we compare our numerical results with the lattice simulations made in Ref.~\cite{adshead}. Figure~\ref{EMratio} shows the time evolution of the ratio $\rho_{\rm EM}/\rho_{\rm tot}$ from the onset of inflation to the end of reheating for a wide range of values for $\alpha$. Note that inflation ends at $\eta=0$ when $t=t_{\rm end}$. Also,
before inflation the time is reckoned by e-foldings $\Delta N \equiv N - N(t_{\rm end})$ and after inflation it is conformal time defined by $\eta=\int_{t_{\rm end}}^t dt'/a(t')$ with rescaled $a(t_{\rm end})=1$. In Ref.~\cite{adshead}, the authors simulated the preheating and reheating processes by tracing the evolution of $\rho_{\rm EM}/\rho_{\rm tot}$ with time after the end of inflation ($\eta>0$) for $\alpha$ ranging from $7$ to $13$. We compare our results for the ratios to those in Ref.~\cite{adshead} with the common input parameters and the same time ranges. We find that overall the two results agree with each other quite well while there are differences in detailed sub-structures. 
Here we are mainly concerned with the inflaton dynamics as well as the gauge particle production near the end of inflation. It is worthy to further scrutinize the preheating and reheating processes and compare the two methods in more details.

\begin{figure}[htp]
\centering
\includegraphics[width=.9\textwidth]{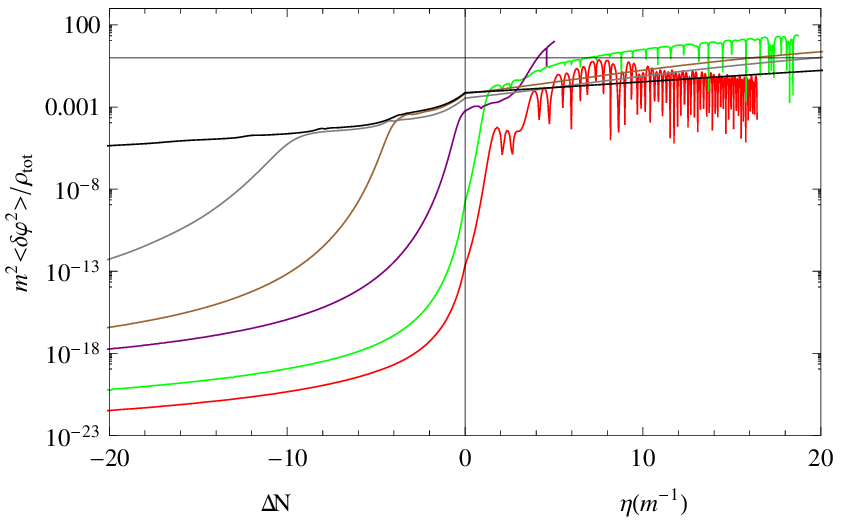}
\caption{Time evolution of the ratio of the inflaton fluctuations, $m^2\langle\delta\varphi^2\rangle$, and the total energy density, $\rho_{\rm tot}\equiv 3M_p^2 H^2$. The lines shown in the figure, in order from bottom to top, are for $\alpha=7,9,14,16,23,32$, respectively.}
\label{phiratio}
\end{figure}

To check the consistency condition for linearity, we have calculated $\langle\delta\varphi^2\rangle$ using Eq.~(\ref{f_powersp}) and plotted the time evolution of the ratio $m^2\langle\delta\varphi^2\rangle/\rho_{\rm tot}$ in Fig.~\ref{phiratio}. The results in the figure are consistent with the large coupling approximation for $\langle\delta\varphi^2\rangle$~(\ref{largealpha}). It seems that in some of the curves the ratio exceeds unity and violates the consistency condition. In fact, before the ratio reaches unity, the reheating process has already finished as shown in Fig.~\ref{EMratio}. For instance, the ratio for the curve with $\alpha=9$ reaches unity at the time $\eta\simeq 7$ while Fig.~\ref{EMratio} shows that the reheating for this case is finished at $\eta\simeq 5$.

\section{Conclusions}

We have constructed a numerical code for computing self-consistently the motion of inflaton, the particle production, and the density perturbation in the pseudoscalar inflation model with the axion-photon-like coupling, $\alpha\varphi F \tilde{F}/(4M_p)$,
where $M_p$ is the reduced Planck mass. We have used a quadratic inflation as a working example and investigated the effects of the coupling for $\alpha<32$. This model is unique in the sense that the photon field receives an effective mass that is proportional to the inflaton velocity.
In particular, that one helicity state of the photon obtains a negative mass leads to a spinoidal instability.
As a consequence, a wide spectrum of photons is produced during inflation efficiently enough
to form a thermal bath that leads to a reheating temperature,
$T_{\rm re} \sim 10^{-6} \alpha M_p$, which is much higher than the perturbative reheating temperature,
$T_{\rm per}\sim 10^{-10} \alpha M_p$. Furthermore, we have calculated the density power spectrum induced by the backreaction of the photon production on the inflaton fluctuations. Using the bound on the density power spectrum coming from the formation of primordial black holes, we have derived a more stringent upper limit $\alpha<20$. In addition, a strong coupling can produce high density spikes in the density power spectrum near the end of inflation that may form primordial black holes.

\label{conclusion}

\begin{acknowledgments}
This work was supported in part by the National
Science Council, Taiwan, ROC under the Grant No.
NSC101-2112-M-001-010-MY3 (K.W.N.) and the Office of Research and Development,
National Taiwan Normal University, Taiwan, ROC under the Grant No. 102A05 (W.L.).
\end{acknowledgments}

\end{document}